\documentstyle[aps,epsfig,manuscript,multicol]{revtex}
\begin{document}
\newcommand{\beq}{\begin{equation}}
\newcommand{\eeq}{\end{equation}}
\newcommand{\sg}{\sigma_p}
\newcommand{\sm}{\sigma_c}
\newcommand{\lf}{\langle L_f \rangle}
\newcommand{\la}{\langle L_a \rangle}
\newcommand{\tf}{{\rm Tr}_f}
\newcommand{\nt}{N_\tau}
\newcommand{\ns}{N_\sigma}
\newcommand{\f}{\beta_f}
\newcommand{\ba}{\beta_a}
\newcommand{\bv}{\beta_v}
\newcommand{\bt}{\beta}
\newcommand{\bvc}{\beta_{vc}}
\newcommand{\lm}{$\lambda$}
\renewcommand{\baselinestretch}{1.0}
\title{Amorphization of Vortex Matter and Reentrant Peak Effect in YBa$_2$Cu$_3$O$_{7-\delta}$}
\author{D. Pal$^1$, D. Dasgupta$^2$, B. K. Sarma$^2$, S. Bhattacharya$^{1,3}$
S. Ramakrishnan$^{1,}$\footnote{E-mail:ramky@tifr.res.in} and A. K. Grover$^{1,}$\footnote{E-mail:grover@tifr.res.in}}
\address{$^1$ Department of Condensed Matter Physics and Materials
Science, Tata Institute of Fundamental Research, Mumbai-400005, India\\
$^2$ Department of Physics, University of Wisconsin, Milwaukee, WI-53201\\
$^3$ NEC Research Institute, 4 Independence Way, Princeton, NJ-08540}
\maketitle
\smallskip
\begin{abstract}
The peak effect (PE) has been observed in a twinned crystal
of YBa$_2$Cu$_3$O$_{7-\delta}$ for H$\parallel$c in the
low field range, close to the zero field superconducting transition temperature (T$_c$(0)) . A sharp depinning transition succeeds the peak temperature T$_p$ of the PE. The PE phenomenon broadens and its internal structure smoothens out as the field is increased or decreased beyond the interval between 250 Oe and 1000 Oe. Moreover, the PE could not be observed above 10 kOe and below 20 Oe. The locus of
the T$_p$(H) values shows a reentrant characteristic with a  nose like
feature located at T$_p$(H)/T$_c$(0)$\approx$0.99 and H$\approx$100 Oe (where
the FLL constant a$_0$$\approx$penetration depth $\lambda$). The
upper part of the PE curve (0.5 kOe$<$H$<$10 kOe) can be fitted to a melting
scenario with the Lindemann number c$_L$$\approx$0.25. The vortex
phase diagram near T$_c$(0) determined from the characteristic features of the PE in YBa$_2$Cu$_3$O$_{7-\delta}$(H$\parallel$c) bears close resemblance to that in the 2H-NbSe$_2$ system, in which a reentrant PE had been observed earlier.\\

\centerline{PACS: 74.60 Ge, 64.70 Dy, 74.25 Dw.}
\end{abstract}
\begin{multicols}{2}
\normalsize
\section{Introduction}
\label{sec:INTRO}
The advent of high T$_c$ superconductors [HTSC] provided a new impetus \cite{r1} to studies on phase transformations in the flux line lattices (FLL) of type-II superconductors in general and they remain a subject of considerable current interest. The competition amongst the elasticity of vortex medium, the pinning of vortex lines and the fluctuation effects of the thermal energy is expected to yield an ordered vortex solid (presumably a dislocation free Bragg glass phase \cite{r2}), a plastically deformed vortex solid ( presumably a vortex glass phase with proliferation of dislocations) and a pinned as well as an unpinned liquid phases in different parts of the field-temperature (H,T) phase space \cite{r1,r2,r3,r4,r5,r6}. In the HTSC cuprate systems, thermal fluctuations are large and in the high temperature part of the (H,T) space (i.e., for $t=\frac{T}{T_{c}(0)}$  $\geq$ 0.9, where T$_c$(0) is the superconducting transition temperature in zero field), a first order melting transition occurs between an ordered solid and a nearly pinning free vortex liquid state \cite{r7,r8,r9}. This transition  in the most investigated YBa$_2$Cu$_3$O$_{7-\delta}$ (YBCO) system is located \cite{r10,r11,r12,r13} at the upper edge of the peak effect (PE) phenomenon in the critical current density J$_c$ for fields of the order of a few tesla. At lower temperatures (i.e., $t < 0.9$) and higher fields, where thermal fluctuations are weaker, a `fishtail' or `second magnetization peak' anomaly occurs \cite{r14} presumably across a $\itshape{transformation}$ between a weakly pinned solid and a stronger pinned solid \cite{r2,r3}. 

The detailed characteristics of the evolution of the anomalous variation in J$_c$ across the (H,T) region of the second magnetization peak and that of the conventional peak effect are far from being understood in the context of a variety of HTSC systems \cite{r14,r15,r16,r17,r18}. In contrast, in the conventional low T$_c$ superconductors (LTSC) with considerably smaller (but often not negligible) thermal fluctuations, a PE near the normal state boundary (H$_{c2}$ line) is a common occurrence over the entire field-temperature regime \cite{r19,r20,r21,r22,r23,r24}. Various attempts \cite{r25,r26,r27,r28} to gain an understanding of the anomalous variations in J$_c$ attribute this phenomenon to a softening of the (weakly pinned) ordered lattice and a consequent transformation into a more strongly pinned amorphous solid or a pinned liquid phase \cite{r29,r30}. Disentangling the effects of the various possible sources of pinning and that of the thermal fluctuations has however remained difficult \cite{r19,r26,r27,r28}. Despite this, there is a widespread acceptance ( supported by evidence from microscopic $\mu$SR studies as well \cite{r31,r32} ) that the anomalous (sharp) variations in J$_c$ signal a change in the state (i.e., in terms of spatial and temporal correlations) of the vortex matter. In the collective pinning framework due to Larkin and Ovchinnikov \cite{r33}, J$_c$ relates inversely to the correlation volume V$_c$ of the Larkin \cite{r34} domain as,
\begin{equation} 
$$ J_cH = \sqrt{n_p<f_p^2>\over{V_c}},  $$
\end{equation} 
 where n$_p$ is the density of pinning centers and f$_p$ is the elementary pinning interaction. Furthermore, in the context of LTSC systems, it is now known \cite{r35} that upon increasing the effective pinning either externally by enhancing the quenched random inhomogeneities in the atomic lattice or by changing the magnetic field in a given sample, the PE often develops  internal structure. This has led to suggestions \cite{r22,r24,r35} of a richer and more complex transformations occurring in a stepwise manner between an ordered ( quasi-lattice) phase  and a fully amorphous one. Even more interestingly, the low field part of the phase diagrams \cite{r37,r38} in the weakly pinned samples of an archetypal LTSC system \cite{r19,r36} 2H-NbSe$_2$  show a variety of novel features, including a reentrance into a disordered phase which is qualitatively similar to a reentrant glass / reentrant pinned liquid, postulated theoretically \cite{r1,r3,r39}.

In this paper, we report on the search and the identification of the anomalous variations in J$_c$ via the ac magnetization measurements in a twinned crystal (T$_c$(0) $\approx$ 93.3 K) of the  YBCO system in the rarely reported low field-high temperature part of the vortex phase diagram. In the YBCO system, the presence of twin boundaries is considered \cite{r10,r17,r27} to facilitate the occurrence (and detection) of the PE phenomenon. The twinned crystal piece (dimensions $\sim$ 0.5$\times$0.5$\times$0.04~mm$^3$ and mass $\sim$ 700 $\mu$g) chosen for the measurements being reported here is obtained by flux growth technique, along with  detwinned YBCO crystals \cite{r40} on which the first order melting transition was 
observed via differential calorimetry \cite{r9}. In this twinned crystal, we are able to detect the PE phenomenon over a wide a field range (20 Oe to 10 kOe),
the lower limit being much lower than any previous study. The evolution 
in the characteristic of the PE is similar to that  in the weakly pinned samples of a variety of LTSC systems \cite{r22,r24,r37,r38}, but some significant differences exist as well.
\section{EXPERIMENTAL DETAILS} 
\label{sec:EXPT}
The ac magnetization measurements were performed using a well shielded home built ac susceptometer \cite{r41}. The dc field (co-axial with ac field) was kept  parallel to the c-axis of the YBCO crystals. Both twinned and detwinned samples 
were investigated but PE was detected only in the former. Data were taken in both the field-cooled and the zero field-cooled (ZFC) modes. Most of the data shown are in the former case, where the PE is more conspicuous.

\section{RESULTS AND ANALYSIS} 
\label{sec:RD}
\subsection{Location of the Peak Effect and elucidation of its characteristic  features}
\label{sec: lOPE}
Figures 1 and 2 present a collation of the temperature dependent in-phase ac susceptibility data recorded with h$_{ac}$ of 0.5 Oe (r.m.s.) and at the frequency 211~Hz in the field ranges where anomalous variations in J$_c$ could be
identified. In zero field, the normalized 4$\pi\chi\prime$ has a value -1 (c.g.s. units) in the superconducting state and it crashes sharply towards the zero value (the transition width (10-90$\%$), $\Delta$T$_c$(0)$\approx$ 0.8 K \cite{r42}) in a featureless manner (see inset of Fig.1(a)). We now focus on the $\chi\prime(T)$ responses shown in the Figs.1(a) through 1(f) as H increases from 20 Oe to 250 Oe. In H=20 Oe (where FLL a$_0$ $\approx$ 1.1$\times$10$^4$\AA), $\chi\prime(T)$ is still featureless, though the superconducting transition is broader. Figures 1(b) and 1(c) show that as H increases from 40 Oe ( a$_0$ $\approx$ 8$\times$10$^3$\AA) to 170 Oe (a$_0$ $\approx$ 4$\times$10$^3$\AA), a characteristic feature reminiscent of the PE develops across the temperature region  marked by a pair of arrows at T$_{pl}$ and T$_p$ in the respective panels. The shape of the $\chi\prime(T)$ curve at H=250 Oe (a$_0$ $\approx$ 3.3$\times$10$^3$\AA) in the inset of Fig.1(f) identifies the PE phenomenon.\\

 Within the Bean's  critical state model  prescription of the magnetization of irreversible superconductors \cite{r43}, the shielding response $\chi\prime(T)$ can be approximated as \cite{r44}:
\begin{equation}
$$  \chi\prime(T) \sim -1 + \alpha \frac{h_{ac}}{J_c}~;~for~h_{ac}~<< H^{\ast},  $$
\end{equation}
\begin{equation}
$$  \chi\prime(T) \sim - \beta \frac{J_c}{h_{ac}}~;~for~h_{ac}   > H^{\ast},  $$
\end{equation}
where $\alpha$ and $\beta$ are geometry and size dependent factors, H$^{\ast}$ is the (threshold) parametric field at which the magnetic field penetrates the center of the sample. Equations 2 and 3 imply that the temperature variation in $\chi\prime$ is governed by the temperature variation of J$_c$ in a given H. Thus, the observed minimum in $\chi\prime(T)$ in the inset of Fig.1(f) marks
a peak in J$_c$, i.e., the PE, which is characterized by two temperatures,
the onset T$_{pl}$ and the peak at T$_p$.
The vanishing of J$_c$ above T$_p$, i.e., the depinning of the vortex state
is characterized by the sharp increase in $\chi\prime$ to its normal state
value. A depinning temperature  T$_{dp}$ (notionally the mid point of
collapse in J$_c$) is obtained from the peak in d$\chi\prime(T)$/dT, examples of which are shown in the insets of Fig.1(b) and Fig.2(d). 

The other noteworthy features contained in Fig.1 and Fig. 2 could be
summarized as follows. \\
{\it (1) The PE is sharpest around an intermediate field of $\sim$1000 Oe, it becomes a broad anomaly for both higher and lower fields, and cannot be detected
above 10 kOe and below 40 Oe. \\
(2) Characteristic structure (e.g., two well resolved minima instead of a single
composite minimum in $\chi\prime$), very similar to what has been seen in LTSC systems, appears in the intermediate field regime as evident  in Figs.1(f), 2(a)
and 2(b). Such a structure has been reported in an other study \cite{r12},
but in higher fields than is shown here. \\
(3) Depinning phenomenon across T$_{dp}$ is a sharp anomaly. A simple fit of the
form: $4 \pi \chi\prime=a~+~T/\Delta T_{dp},$~yields a characteristic
width $\Delta$T$_{dp}$, whose field dependence is shown in the inset of Fig.2(b). Interestingly, $\Delta$T$_{dp}$(H)
is smallest for H$\approx$1000 Oe and increases for both increasing and decreasing fields, as shown in the inset of Fig.2(b). The $\Delta$T$_{dp}$(H) at 1~kOe is 
smaller than its value even at zero field (see the encircled data point), the latter is very close to the $\Delta$T$_c$(0), which is shown in the inset of Fig.1(a).} 
\subsection{Observation of the Differential Paramagnetic Effect}
\label{sec: DPE}
 
Figures 3(a) and 3(b) show the $\chi\prime(T)$ values measured with a higher h$_{ac}$ of 2~Oe (r.m.s.) at a frequency of 211Hz for the vortex states obtained in 320 Oe and 20 Oe, respectively. The PE in the main panel of Fig.3(a) is pronounced \cite{r45}. A significant new feature to note, however, is that  $\chi\prime$ makes a sharp transition from diamagnetic values to {\it paramagnetic} values across the depinning temperature T$_{dp}$. The inset in Fig.3(a) reveals  a small paramagnetic peak above the temperature T$_{irr}$ (only slightly greater than the so designated T$_{dp}$), the ubiquitous irreversibility temperature. The paramagnetic $\chi\prime$ response returns to the background level on crossing over to the normal state at T$_c$(H). An identical paramagnetic peak located at the edge of the depinning transition has earlier been reported in LTSC systems, such as, CeRu$_2$ \cite{r45} and  2H-NbSe$_2$ \cite{r46} and identified with the {\it differential paramagnetic effect} (DPE) \cite{r47}. Between T$_{irr}$(H) and T$_c$(H), the vortex state is reversible, for which dM/dH is positive, i.e., diamagnetic dc magnetization decreases as H increases, implying, a DPE. 
The paramagnetic peaks in the $\chi\prime(T)$ data in single crystals of YBCO and BSCCO had been reported in other studies as well \cite{r48,r49,r50}. 
Morozov et al \cite{r48} had reported a sharp paramagnetic peak in a single crystal of Bi$_2$Sr$_2$CaCu$_2$O$_8$ (BSCCO) via local micro-Hall 
sensors at temperatures where a step change in equilibrium magnetization (M$_{eq}$) due to FLL melting is expected. However, no peak effect has been
reported in these studies.
The question, as to whether the sharp onset of the DPE peak marks the step change in M$_{eq}$ at the FLL melting transition \cite{r48} and/or it reflects the depinning transition located at the upper edge of the PE phenomenon, requires more detailed investigations. We note, however, that Ishida {\it et al} \cite{r12} and Ravikumar et al \cite{r51} have presented evidence of a step change in M$_{eq}$ across the PE region in the low field-high temperature part of the (H,T) phase space in single crystals of YBCO and 2H-NbSe$_2$, respectively. Thus, the sharp transition from the PE peak to a DPE peak in weakly pinned systems (as in Fig.3(a)) may indeed mark the transition from a pinned to an unpinned state of
the vortex matter.

Furthermore, the Fig.3(b) shows yet another effect of a larger ac amplitude:
a residual PE now visible in contrast with the low amplitude data in
Fig.1(a). It is possible that a larger ac signal anneals \cite{r52} the dilute
lattice thereby making the amorphization more easily detectable.
The possible annealing effect of a larger ac field needs further investigations.
Studies on samples of LTSC had earlier revealed \cite{r45} that specific details 
of the structure in the $\chi\prime(T)$ curve across the PE depend on the 
amplitude of the ac field.
\subsection{History effects in $\chi\prime(T)$ behavior}
\label{sec: HISTORY}
The history effects pertain to the dependence of $\chi\prime(H,T)$ on the path followed in reaching a given H and T. A simple way to explore them is to examine the difference in $\chi\prime(T)$ between ZFC and FC modes. Fig.4 displays the $\chi\prime(T)$ data in h$_{ac}$ of 2 Oe (r.m.s.) at H = 500 Oe ($\Arrowvert$c) (i.e., when the PE peak is well recognizable) for both ZFC and FC modes. The inset in Fig.4 shows the data across the PE region on an expanded scale. Note first that above T$_p$, $\chi\prime(T)$ behavior is the same for both ZFC and FC modes. However, prior to the entry into the PE region (i.e., for $T < T_{pl}$), $\arrowvert\chi\prime_{FC}\arrowvert > \arrowvert\chi\prime_{ZFC}\arrowvert $ . As per eqn.(2) and eqn.(3), this inequality would translate as  J$_c^{FC}$ $\leq$ J$_c^{ZFC}$ for $T < T_p$. In terms of Larkin-Ovchinnikov scenario (cf. eqn.(1)), this implies that the correlation volume V$_c$ in the FC state in larger than that in the ZFC state. This inference agrees with the direct observation of better order for the vortex states  (mostly at low fields) prepared in the FC manner as compared to those prepared in the ZFC manner in crystals of YBCO \cite{r53} as well as in other high T$_c$ cuprates \cite{r54}. However, the above inequality is in complete contrast to the situation in weakly pinned samples of LTSC systems on which history dependent J$_c$ data have been reported for long \cite{r55,r56,r57,r58,r22,r24}.

In transport studies of single crystals of niobium,
Steingart, Putz and Kramer \cite{r55} had noted  the inequality,
\begin{equation} 
$$J_c^{FC}(H) > J_c^{rev}(H) >J_c^{ZFC}(H),$$ 
\end{equation}
where $J_c^{rev}(H)$ is the current density while reversing the field from above H$_{c2}$ to a given H in an isothermal scan. In recent years, the history effects and metastability (supercooling, etc.) has received new impetus \cite{r57}. The observations that (i)$J_c^{FC}(H) > J_c^{ZFC}(H)$  for $H < H_p$ and (ii)$J_c^{FC}(H) \approx J_c^{ZFC}(H)$ for $H > H_p$ in LTSC systems have led to the suggestion \cite{r22} that FC states attempt to freeze in (i.e., supercool) the amorphous correlations present at (and above) the peak position of the PE. A FC state is therefore more strongly pinned than the corresponding ZFC state. 
The ZFC phase, in contrast, is considered to be prepared by exposing the weakly pinned superconducting sample to an applied field which generates vortices that enter the sample at high velocities \cite{r59} and are able to overcome the effects of pinning centers to eventually explore the well ordered stable state of the system. It may be speculated here that while field cooling in the YBCO case, the temperature interval between the irreversibility temperature T$_{irr}$ and the peak temperature T$_p$ is so narrow (cf. Fig.3(a)) that the sample fails to explore the (equilibrium) amorphous phase during a very fast cool down. The vortex density in the sample is uniform in the FC state, whereas in the ZFC case, in particular at low fields, the strong pinning effects near the edges in the
platlet shaped samples of YBCO result in the vortex density being non-uniform across the cross-section of the sample. Large thermal energy would help overcome the effects of pinning centers in both cases, however, the FC state being more (macroscopically) uniform eventually produces a better ordered FLL with a larger correlation volume V$_c$ of the Larkin domain (than that in the ZFC case). We
note that Kokkaliaris et al \cite{r18} have recently reported that in an untwinned YBCO crystal, which displays the phenomenon of a broad second magnetization peak (to be designated as H$_p^s$) at high fields and lower temperatures, $J_c^{ZFC}(H) < J_c^{rev}(H)$  for $H < H_p^s$. Such an inequality is indeed consistent with the behavior in LTSC systems (cf. eqn.(4)).
The role that the width of the PE region could play in deciding the nature of the inequality between J$_c^{FC}$ and J$_c^{ZFC}$ in LTSC/HTSC systems remains to be further investigated. 

It is to be noted that regardless of the sign of the history dependence, in
all cases reported so far, the T$_p$ provides an upper limit of the history
dependence. Thus, in analogy with disordered magnets such as spin glasses,
T$_p$ marks a characteristic ``transition" temperature above which the system
is disordered in ``equilibrium".

\subsection{Construction of the vortex phase diagram and its comparison with earlier reports}
\label{sec:PHASE}
Figure 5 shows a collation of the  data corresponding to the PE region marked by the onset (T$_{pl}$) and the peak (T$_{p}$) positions along with the depinning 
(T$_{dp}$) and the superconducting transition (T$_c$) temperatures over the (H,T) region, where the PE phenomenon could be identified. In view of the observation that the sharp depinning transition (i.e., the irreversibility temperature) immediately succeeds the peak of the PE, the T$_p(H)$ line in Fig.5 runs parallel to the T$_{dp}(H)$ line. Several reports \cite{r10,r11,r12,r13,r17} in the twinned and untwinned crystals of YBCO provide  evidence that the T$_p(H)$ values are located either in close proximity to the melting temperatures (as determined from the transport studies \cite{r10,r11,r17}) of the underlying pristine FLL or they coincide with the temperatures of a step change in the equilibrium magnetization \cite{r12,r13,r17}. One could draw the loci of different features of the PE phenomenon seen in transport \cite{r11,r17} and ac susceptibility measurements \cite{r12,r13,r17}, however, they all seem to indicate that H-T lines so drawn would conform to the power-law FLL melting relationship \cite{r1}:
\begin{equation}
$$ H_m=H_0(1-\frac{T}{T_c(0)})^n   ,   $$
\end{equation}
where the prefactor H$_0$ $\sim$ 10$^6$ Oe and n $\approx$ 1 to 2  \cite{r9,r11,r13,r17,r18}. Most of the data on the PE curve or the melting line or the irreversibility line ( in YBCO ) in the literature, which have been found to conform to the power law relationship, are at fields larger than 1 kOe. A broken line satisfying the T$_{dp}(H)$ data points for H $>$ 0.25 kOe in Fig.5 
attests to the efficacy of the power law behavior. We, however, find that in the low field region (H$<$0.5 kOe), the deviations from the power law relationship 
(H$\sim$(1-t)$^n$) set in a significant manner for both the curves T$_{pl}(H)$ and T$_p(H)$.

We recall that a sudden collapse of the irreversibility line between 1 kOe and 0.2 kOe in the twinned and the untwinned crystals of YBCO was reported by Krusin-Elbaum et al \cite{r60} from high frequency ac susceptibility measurements. However, in their data, there is no evidence of the occurrence of a PE. Below 200 Oe, the irreversibility line of Krusin-Elbaum et al \cite{r60} can be seen to proceed towards the T$_c(0)$ value in a power-law manner once again, but with a different value of the exponent n. Following Ref.60, we show in Fig.6, the plot of log~H versus log~(1-t) for the T$_{dp}$ line in our sample of YBCO. Note that the data for 6000 Oe $\leq$ H $\leq$ 250 Oe and for 70 Oe $\leq$ H $\leq$ 20 Oe could be considered to conform to the power law relationship albeit with different values of the exponent n in the two intervals \cite{r60}. In the high field region ( H $>$ 0.25 kOe), both the exponent n$\approx$2 and the prefactor $H_0 \approx 7.9\times10^6~Oe$ are comparable to similar values (n$\approx$2 and $H_0 \approx 6\times10^6~Oe$) found by Nishizaki et al \cite{r17} in a high quality twinned crystal of YBCO. In between the high field and low field regions, i.e., between 250 Oe and 70 Oe, T$_{dp}$ curve shows a nose like turnaround feature. Such a feature appears accentuated in T$_p(H)$ and T$_{pl}(H)$ curves, as can be clearly viewed via a replot of the vortex phase diagram in a semi-log manner as shown in the Fig.7. In order to comprehend the genesis of the turnaround in T$_p(H)$ curve around 100 Oe, we refer to the $\chi\prime(T)$ curves displayed in Figs.1(b) to 1(f). These figures show that the PE develops gradually, followed by
a progressively rapid depinning immediately thereafter (cf. data in the inset of Fig.2(b)). A careful examination of the $\chi\prime(T)$ curves in Figs.1(b), 1(c) and 1(d) at H = 40 Oe, 70 Oe and 120 Oe shows how the sharpening of the PE features results in an increase in T$_p(H)$ ( and T$_{dp}(H)$) values with 
increasing  H. Above about 150 Oe, T$_p(H)$ values start to show the usual decrease with increase in H. The shape of the T$_p(H)$ curve in Fig.7 is reminiscent of the reentrant characteristic in the PE curve noted first by Ghosh et al \cite{r37} in  NbSe$_2$. 

We further note that the broadening and eventual disappearance of the PE phenomenon at fields less than 30 Oe is also similar to the behavior in NbSe$_2$. If we assume that the PE curve(s) separates the ordered and disordered phases of vortex matter then the absence of the PE at low fields 
($<$30 Oe, where a$_0$ $>$ 1$\mu$) amounts to the absence of an ordered phase, when the dilute vortices are individually pinned by the quenched random disorder in the atomic lattice. In analogy with the results in NbSe$_2$ \cite{r37} and other LTSC systems \cite{r22,r24}, we have chosen to label the vortex phase between the higher field upper portion of the T$_p(H)$ curve and the T$_{dp}$ line as the pinned amorphous (analogous to the pinned liquid) phase. The (H,T) region below the reentrant leg of the T$_p(H)$ curve at low fields has been termed as the `reentrant disordered' ( and filled with dotted lines) in the  Fig.7. The reentrant disordered and pinned amorphous regions would overlap near the turnaround in the T$_p(H)$ curve. At temperatures above the turnaround feature
(t $>$ 0.99), the vortex state remains disordered at all fields. However, below this temperature (i.e., t$<$0.98) an ordered vortex phase (presumably akin to an elastic/Bragg glass) exists between the high field upper portion of T$_{pl}(H)$ curve and the `reentrant disordered' phase \cite{r46}. The vortex phase between T$_{pl}(H)$ and T$_p(H)$ curves is expected to be  plastically deformed \cite{r11,r19} and has therefore been termed as plastic glass \cite{r22,r38,r46}, detailed theoretical understanding of which is still lacking. The vortex phase diagram in YBCO (in Fig.5/Fig.7) thus shows a close resemblance with the generic phase diagram for a weakly pinned superconductor drawn by Banerjee {\it et al} \cite{r38,r46} on the basis of characteristics of the peak effect in the NbSe$_2$ system.\\

It is instructive to dwell a little more on the similarities in the present YBCO crystal and the 2H-NbSe$_2$ sample studied by Ghosh et al \cite{r37}. We note that the nose feature in t$_p$ (=$\frac{T_p(H)}{T_c(0)}$) curve occurs at nearly the same (H,t) value of ($\sim$ 100 Oe, $\sim$0.99) and the PE also disappears in both the samples at H$<$30 Oe. Further, the FLL spacing a$_0$ near the nose feature in NbSe$_2$ was comparable to its in-plane penetration depth $\lambda_{ab}$ \cite{r46}. We find the similar correspondence in YBCO; a$_0$ at 100 Oe is 
$\approx$5.0$\times$10$^3$\AA~and the measured value of $\lambda$ is $\approx$ 5.5$\times$10$^3$\AA~\cite{r61} at t$\approx$0.99 in another crystal of YBCO grown by flux growth technique. The upper portion of the t$_p$ curve in 2H-NbSe$_2$ was fitted (see inset in Fig.2 of Ghosh {\it et al} \cite{r37}) to the FLL melting relation with the exponent n=2 and the prefactor $H_0~=~\beta_m(c_l^4/G_i)H_{c2}(0)$,  where the various  symbols have their usual meaning \cite{r1}. It yielded the Lindemann number c$_L$ in NbSe$_2$ as 0.17. Following Ghosh et al \cite{r37}, we plot  H versus (1-t$_p$)$^2$ in YBCO in the inset of Fig.6. A linear fit to the data above 500 Oe in this inset yields $c_L\approx0.25$ (where H$_{c2}(0)$ $\approx$ 150T, $\beta_m = 5.6$, 
$G_i = 10^{-2}$ \cite{r1}) The disappearance of the PE at low fields is rationalized \cite{r37} by invoking the notion that the pinning length L$_c$ and the entanglement length L$_E$ become comparable (cf. eqn.(6.47) of \cite{r1}) at H $\approx$ 30 Oe. Recalling once again the same relation, the L$_c$/L$_E$ is given as, 
$$ \frac{L_c}{L_E}\simeq \frac{\pi\kappa^{2}ln\kappa}{\surd2}(\frac{a_0}{2\pi\lambda(0)})^2(\frac{j_c}
{G_{i}j_{0}})^{\frac{1}{2}}\frac{(1-t)^{\frac{4}{3}}}{t}, \eqno(7) $$

where the various symbols have their usual meaning \cite{r1}. Note, how the 
 L$_c$/L$_E$ scales with the crucial parameters, like, the Ginzburg number 
G$_i$($=(1/2)(k_BT_c/H_c^2\xi^3\epsilon)^2$) and the ratio j$_c$/j$_0$, where j$_c$ and j$_0$ represent the critical current density and the depairing current density, respectively. The G$_i$ in YBCO ($\approx$ 10$^{-2}$) is expected to be two to three orders larger than that in 2H-NbSe$_2$ \cite {r19,r37}, but it is compensated by the extreme smallness of j$_c$/j$_0$ value ( $\sim$ 10$^{-5}$ to 10$^{-6}$) in typically clean crystals of 2H-NbSe$_2$ as compared to that ($\sim$ 10$^{-2}$ - 10$^{-3}$) in a clean YBCO sample \cite{r19}. At fields below 30 Oe, where L$_c$/L$_E$$<$1, the dilute vortex array is expected to be so disordered
that it does not display any anomalous variation in J$_c$(T) corresponding
to an order-disorder transformation in an isofield scan. This order of magnitude estimate for the field value of disorder-order transition in FLL correlations appears consistent with isothermal Bitter decoration data in samples of YBCO \cite{r62} and BSSCO \cite{r63}.
\section{Conclusion}
\label{sec:CON}
In summary, we have reported the occurrence of the peak effect phenomenon for sparse vortex arrays at very low fields as well as for the dense vortex arrays at moderately high fields in a twinned crystal of YBCO. The data  delineate the order-disorder transformation regions at low fields, where the interaction is weaker and the disorder effects of the thermal fluctuations and/or small bundle (or individual) pinning dominate. The new observations at low fields agree  with an earlier report \cite{r60} of a collapse of the irreversibility line in the (H,T) region close to T$_c(0)$ in the context of the YBCO system.
But, in addition, it shows  a close connection with the ``nose" like turnaround feature in the locus of the peak effect in a LTSC 2H-NbSe$_2$ system
\cite{r37}. The notion of the thermal softening of the vortex cores \cite{r64} has been invoked \cite{r60} for YBCO, whereas a connection with the reentrant melting/reentrant disordered phase was proposed for 2H-NbSe$_2$. 
Both ideas imply  a transformation/crossover from an interaction dominated collective pinning behavior to a different regime, where pinning/depinning 
of vortex lines have to be considered in isolation. The observation of the complexity in the PE regime  (at intermediate fields), which comprises the onset of a sudden shrinkage in the correlation volume \cite{r22,r24} and the internal structure, reinforces the notion of a stepwise loss of vortex correlations.  Carruzzo and Yu \cite{r65} have recently theoretically proposed a two step 
softening process of the FLL, in which a first order transition initially transforms an ordered vortex solid to a (defective) soft vortex solid that has smaller, but, still finite shear modulus. The latter solid in turn could undergo another first order transition to the vortex liquid state. Our present results
call for more detailed study on the PE in different systems to explore the
loss in  order of the FLL.
The presence of twin boundaries in YBCO \cite{r10} is usually considered to be detrimental to the observation of a first order melting transition. However, in our case, the presence of twin boundaries has facilitated the observation of the peak effect at low fields. Further, a 
recent study of the first order FLL melting transition on the twinned crystals of NdBa$_2$Cu$_3$O$_7$ \cite{r66} also shows the usefulness of the twinned crystals to explore basic issues in vortex state physics. As a final remark, we recall that 
Grier {\it et al} \cite{r63} and Horiuchi et al \cite{r67} have directly demonstrated how the  disordered dilute vortex arrays undergo  transformation to an ordered phase as the interaction effect increases due to an increase in the vortex density either by an increase in the applied field
\cite{r63,r67} or, more interestingly,  by the collation of the vortices around  a strong pinning center \cite{r67}. Such studies in samples of YBCO and NbSe$_2$, which show reentrant characteristic in the peak effect curve are highly 
desirable. \\
~~\\
\centerline{\bf Acknowledgements}
~~\\
It is a pleasure to acknowledge numerous discussions with Satyajit Banerjee,
Shampa Sarkar, Mahesh Chandran, Nandini Trivedi, G. Ravi Kumar, 
Prasant Mishra, Vinod Sahni and Gautam Menon.
\end{multicols}

\large
\begin{center}
{\bf FIGURE CAPTIONS}
\end{center}
\normalsize
Fig.   1:  Temperature dependence of the normalized values of the in-phase ac susceptibility (4$\pi\chi\prime(T)$) in a twinned single crystal of YBa$_2$Cu$_3$O$_7$ measured with an amplitude (h$_{ac}$) of 0.5 Oe (r.m.s.) and at a frequency of 211 Hz in the dc fields ($\|$c) as indicated in the different panels. The insets in the Figs. 1(a) and 1(f) show the $\chi\prime(T)$ response in (nominal) zero field and 250 Oe, respectively. The onset (T$_{pl}$) and peak (T$_p$) temperatures of the PE have been marked in different panels. The inset panel in Fig.1(b) shows the plot of d$\chi\prime$/dT vs T in H = 40 Oe. The temperature of maximum in d$\chi\prime$/dT marks the mid point of the depinning transition (T$_{dp}$). An unlabelled arrow  in the Fig. 1(a) identifies the temperature across which a remanence of the PE phenomenon can be identified in the $\chi\prime(T)$ curve at H = 20 Oe measured with a higher h$_{ac}$ of 2 Oe (r.m.s.), as shown in the Fig.3(b).

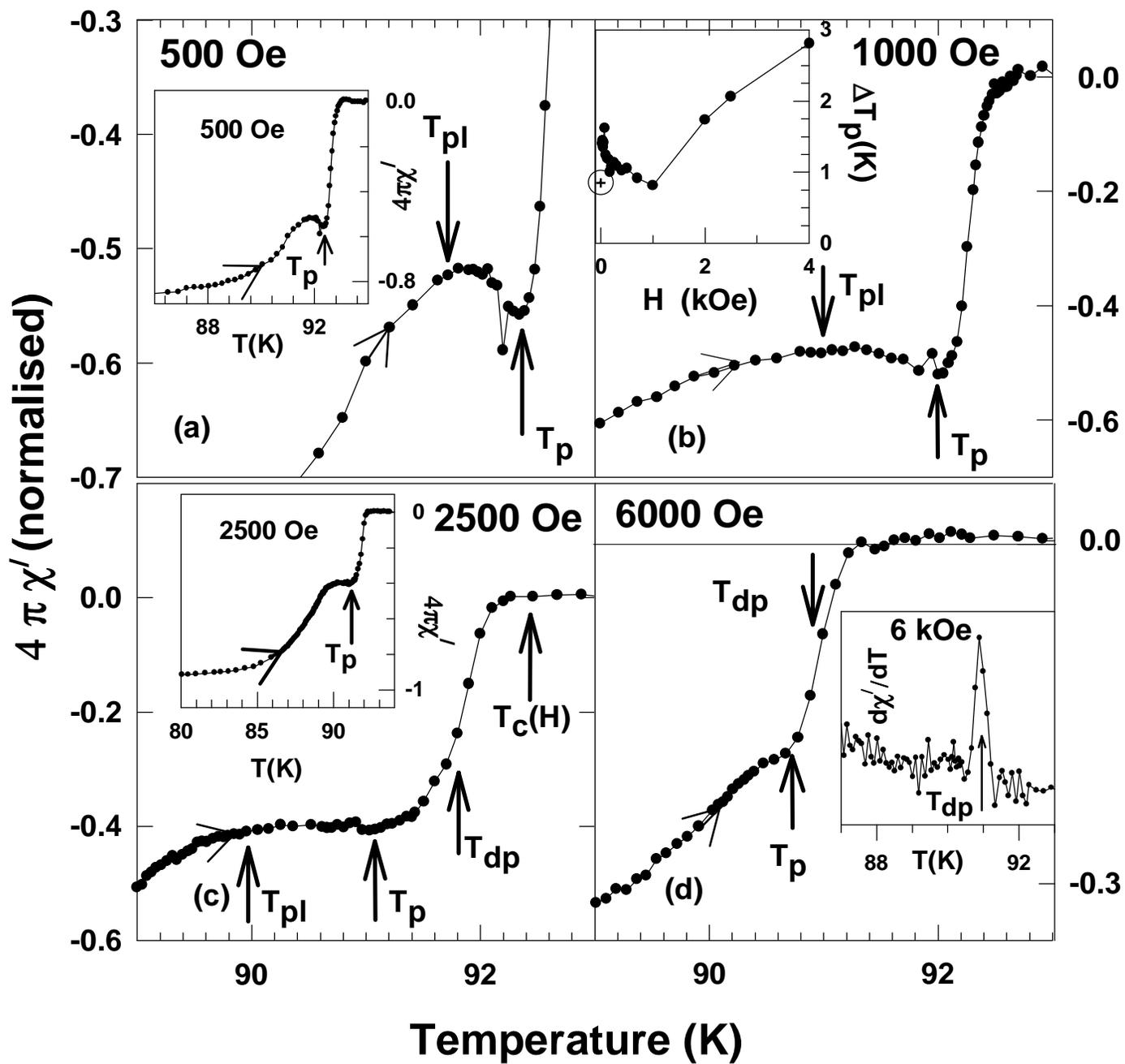
Fig.   2:  Plots of 4$\pi\chi\prime$ vs T in higher dc fields. The inset panels in Fig.2(a) and Fig.2(c) help to view the occurrence of the PE at H = 500 Oe and 2500 Oe, respectively. The inset in Fig.2(d) shows the location of the T$_{dp}$ via a plot of d$\chi\prime$/dT versus T at H = 6000 Oe. The inset in Fig.2(b) shows a plot of the parameter $\Delta$T$_{dp}$ versus field (see text for details).

Fig.   3:  The $\chi\prime(T)$ measured in a h$_{ac}$ of 2 Oe(r.m.s.) and at a frequency of 211 Hz in (a) 320 Oe and (b) 20 Oe. An enlarged plot in the inset of panel (a) shows the presence of the differential paramagnetic effect (DPE), which follows the occurrence of the peak effect. The inset of panel (b) shows an inflection feature in the $\chi\prime(T)$ curve at H = 20 Oe.

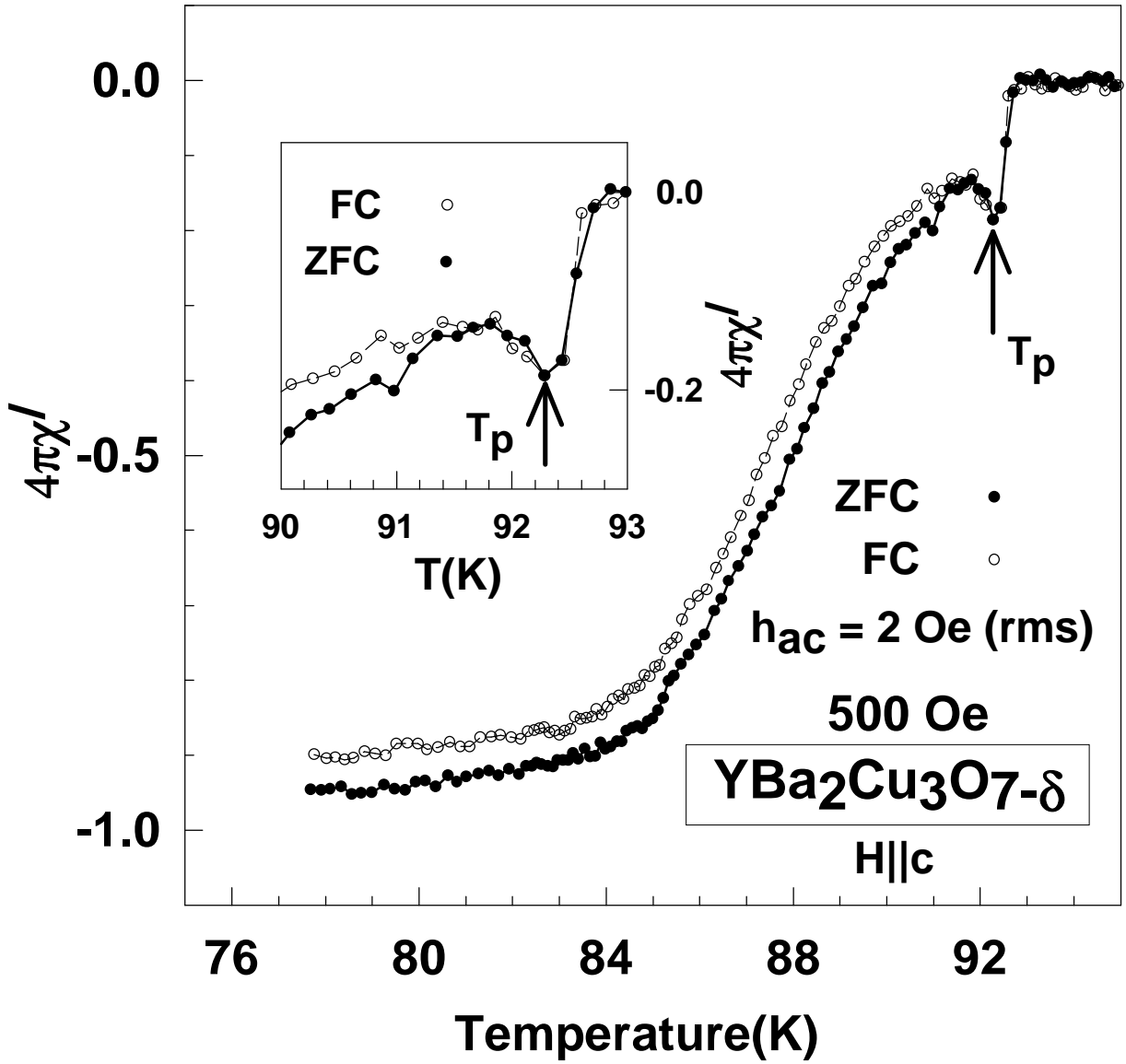
Fig.   4:  Comparison of  the $\chi\prime(T)$ behavior (h$_{ac}$ = 2 Oe (r.m.s.) and f = 211 Hz) for the vortex states prepared at 500 Oe in the zero field cooled (ZFC) and the field cooled (FC) modes. The inset shows an enlarged view of the data across T$_p$.

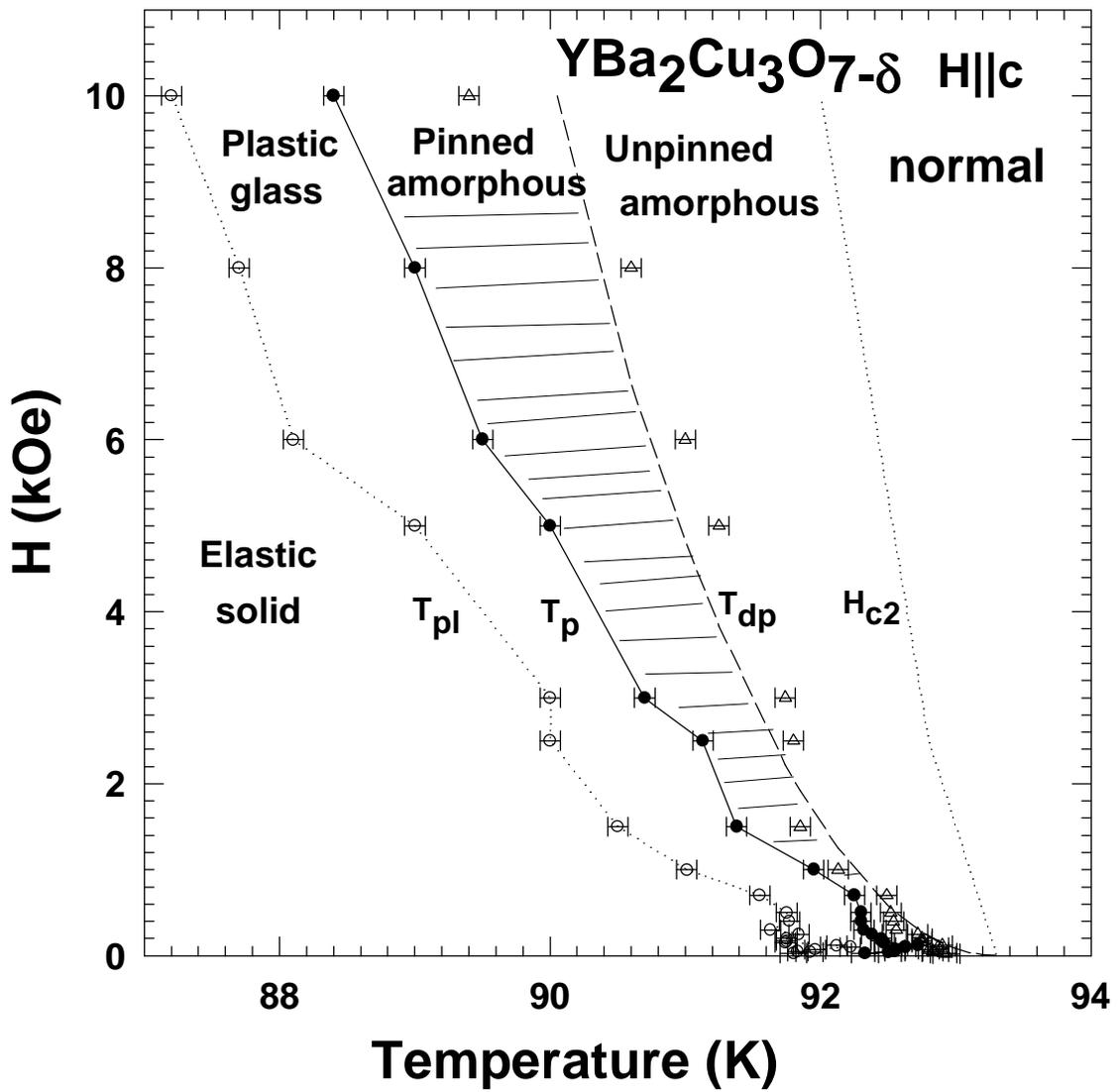
Fig.   5:  The vortex phase diagram in the twinned YBa$_2$Cu$_3$O$_{7-\delta}$ crystal (for H$\|$c) constructed on the basis of the data on the onset (T$_{pl}$), the peak (T$_p$) and the end (i.e., the depinning temperature T$_{dp}$) of the peak effect phenomenon. The schematically drawn normal state boundary (i.e., the dotted H$_{c2}$ line) corresponds to the (nominal) zero field transition temperature of 93.3 K and the T$_c(H)$ values ascertained at high fields (H $>$ 0.25 kOe). The dashed line notionally marks the upper end of the pinned amorphous region and it corresponds to the relationship, H = 7.9$\times$10$^6$(1-$\frac{T}{T_c(0)}$)$^2$ Oe. For a justification of the nomenclature of the different vortex phases, see text.

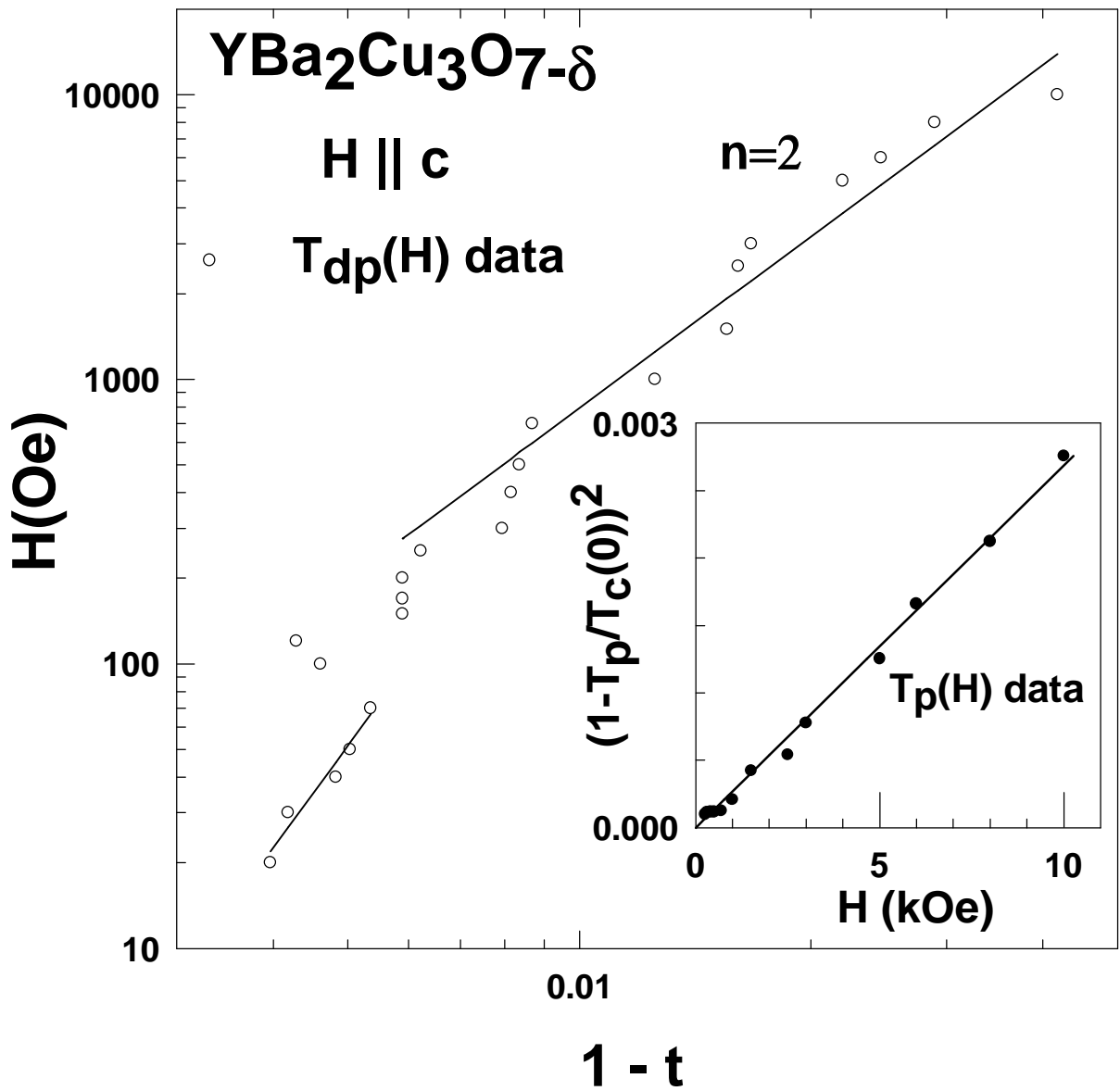
Fig.   6: Log-log plot of H vs (1-t), where $t = \frac{T_{dp}}{T_c(0)}$. The solid lines correspond to the power law [ $H \approx H_0(1-t)^n$ ], above and below the field-temperature region of a collapse (or turn around) in the depinning line. The high field ( H $>$ 0.3 kOe) exponent is n $\approx$ 2 and the prefactor $H_0 \approx 7.9\times10^6 Oe$. The inset shows a plot of H vs (1-t$_p$)$^2$, where $t_p = \frac{T_p}{T_c(0)}$. T$_p(H)$ data yield prefactor H$_0$ $\approx$ 3.7$\times$10$^6$ Oe.

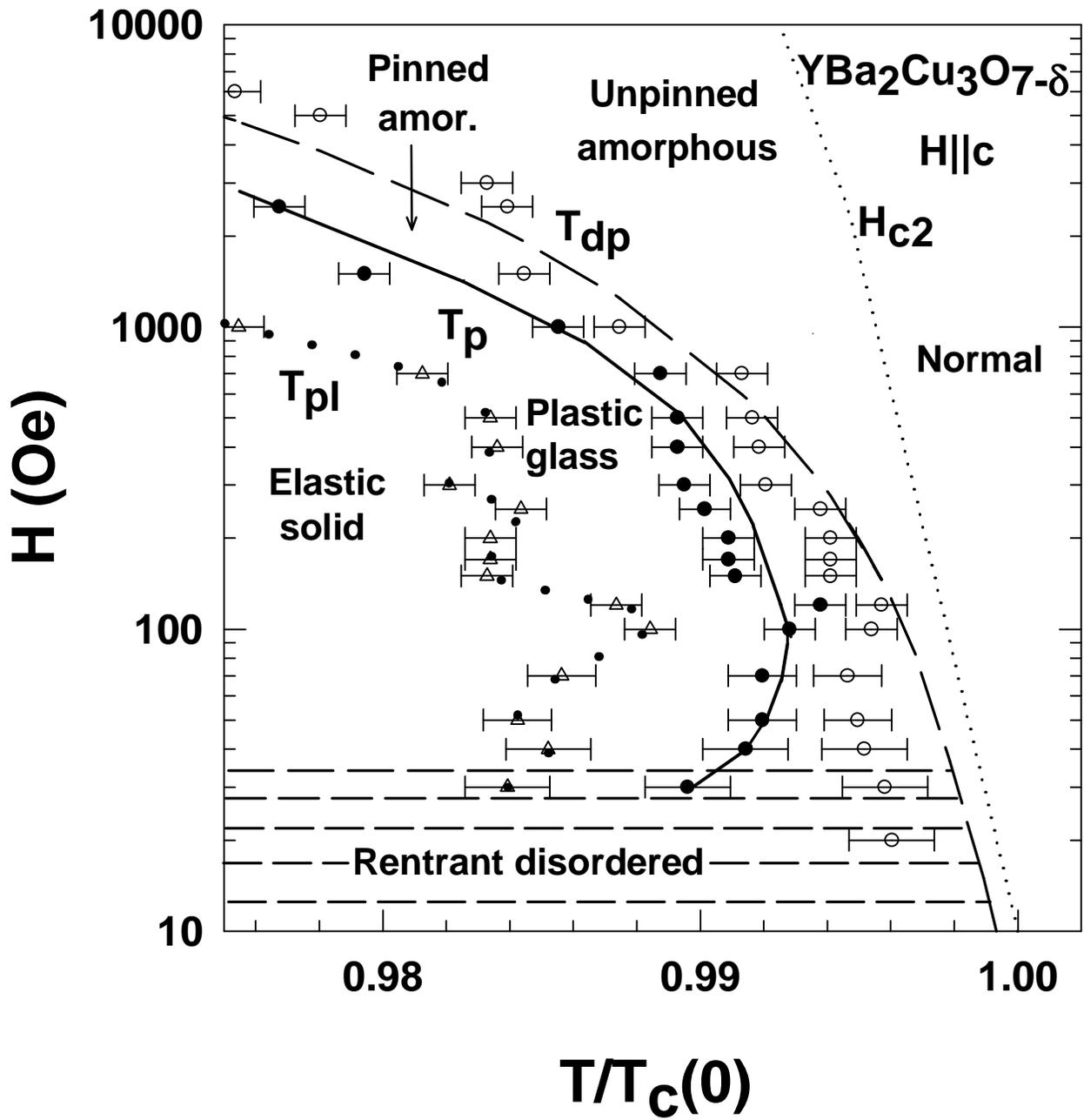
Fig.   7:  The same data as in Fig.5 on a semilog plot illustrating the shape of the  t$_p(H)$ curve at low fields ( H $<$ 300 Oe). As in Fig.5, the dashed line conforms to the power law relation with exponent n $\approx$ 2. The solid line satisfying the t$_p(H)$ data points has been drawn in a free hand manner, it can be seen to run parallel to the dashed line for H $>$ 300 Oe. The dotted line joining $t_{pl}(H)$ data points is a guide to the eye.

\begin{references}
\bibitem{r1} G. Blatter, M. V. Feigel'man, V. B. Geshkenbein, A. I. Larkin and V.  M. Vinokur, Rev. Mod. Phys. {\bf 66}, 1125 (1994).
\bibitem{r2} T. Giamarchi and P. Le Doussal, Phys. Rev. Lett. {\bf 72}, 1530  (1994); Phys. Rev. B {\bf 52}, 1242 (1995).
\bibitem{r3} M. J. P. Gingras and D. A. Huse , Phys. Rev. B {\bf 53}, 15193
 (1996).
\bibitem{r4} D. Ertas and D. R. Nelson, Phys. Rev. B {\bf 53}, 193 (1998).
\bibitem{r5} J. Kierfield, T. Natterman and T. Hwa, Phys. Rev. B {\bf 55}, 626 (1997).
\bibitem{r6} T. Giamarchi and P. Le Doussal, Phys. Rev. B {\bf 55}, 6577 (1997).
\bibitem{r7} H. Safar, P. Gammel, D. A. Huse, D. J. Bishop, W. C. Lee,
J. Giapintzakis and D. M. Ginsberg, Phys. Rev. Lett. {\bf 69}, 824 (1992); 
{\bf 70}, 3800 (1993).
\bibitem{r8} E. Zeldov, D. Majer, M. Konczykowski, V. M. Vinokur
and H. Shtrikman, Nature (London), {\bf 375}, 373 (1995).
\bibitem{r9} A. Schilling, R. A. Fisher, N. E. Phillips, U. Welp,
D. Dasgupta, W. K. Kwok, G. W. Crabtree, Nature (London), {\bf 382}, 791 (1996).
\bibitem{r10} K. Kwok, J. A. Fendrich, C. J. van der Beek and C. W. Crabtree, Phys. Rev. Lett. {\bf 73}, 2614 (1994).
\bibitem{r11} W. Jiang, N. C. Yeh, T.A. Tombrello, A. P. Rice and F. Holtzberg,
 J. Phys. Condensed Matter {\bf 9}, 8085 (1997).
\bibitem{r12} T. Ishida, K. Okuda and H. Asaoka, Phys. Rev. B {\bf 56}, 5128 (1997).
\bibitem{r13} J. Shi, X. S. Ling, D. A. Bonn and W. M. Hardy,
 Phys. Rev. B {\bf 60}, R12593 (1999).
\bibitem{r14} M. Daeumling, J. M. Seuntjens and D. C. Larbalestier, 
Nature (London) (London) {\bf 336}, 332 (1990); K. Deligiannis, P. A. J. de Groot, M. Oussena, S. Pifold, R. H. Langan, R. Gagnon and L. Taillefer, 
Phys. Rev. Lett. {\bf 79}, 2121 (1997).
\bibitem{r15} B. Khaykovich, E. Zeldov, D. Majer, T. W. Li, P. H. Kes and M. Konczykowski, Phys. Rev. Lett. {\bf 76}, 2555 (1996).
\bibitem{r16} H. K\"upfer, Th. Wolf, C. Lessing, A. A. Zhukov, X. Lan‡on, R. Meier-Hirmer, W. Schauer and H. W\"uhl, Phys. Rev. B {\bf 58}, 2886 (1998) and references cited therein.
\bibitem{r17} T. Nishizaki, T. Naito, and N. Kobayashi, Phys. Rev. B, {\bf 58}, 11169 (1998); Phys. Rev. B {\bf 53}, 82 (1996); J. Low Temp. Phys. {\bf 105}, 1183 (1996).
\bibitem{r18} R. M. Langan, S. N. Gordeev, P. A. J. de Groot, A. G. M. Jansen, R. Gagnon and L. Taillefer, Phys. Rev. B {\bf 58}, 14548 (1998); S. Kokkaliaris, D. A. J. de Groot, S. N. Gordeev, A. A. Zhukov, R. Gagnon and L. Taillefer, Phys. Rev. Lett. {\bf 82}, 5116 (1999).
\bibitem{r19} M. J. Higgins and S. Bhattacharya, Physica C {\bf 257}, 232 (1996) and references cited therein.
\bibitem{r20} K. Gloos, R. Modler, H. Schimanski, C. D. Bredl, C. Geibel, F. Steglich, A. I. Buzdin, N. Sato and T. Komatsubara, Phys. Rev. Lett. {\bf70}, 501 (1993); A. Ishiguro, A. Sawada, Y. Inada, J. Kimura, M. Suzuki, N. Sato and T. Komatsubara, J. Phys. Soc. Jpn. {\bf 64}, 378 (1995).
\bibitem{r21} R. Wordenweber and P. H. Kes, Phys. Rev. B {\bf 34}, 594 (1986).
\bibitem{r22} S. S. Banerjee, N. G. Patil,  S. Saha, S. Ramakrishnan, A. K. Grover, S. Bhattacharya, G. Ravikumar, P. K. Mishra, T. V. Chandrasekhar Rao, V. C. Sahni, M. J. Higgins, E. Yamamoto, Y. Haga, M. Hedo, Y. Inada and Y. Onuki, Phys. Rev. B {\bf 58}, 995 (1998).
\bibitem{r23} K. Tenya, S. Yasunami, T. Tayama, H. Amitsuka, T. Sakakibara, M. Hedo, Y. Inada, E. Yamanoto, Y. Haga and Y. Onuki, J. Phys. Soc. Jap. {\bf 68}, 224 (1999) and references citeed  therein.
\bibitem{r24} S. Sarkar, D. Pal, S. S. Banerjee, S. Ramakrishnan, A. K. Grover, C. V. Tomy, G. Ravikumar, P. K. Mishra, V. C. Sahni, G. Balakrishnan, D. Mck. Paul and S. Bhattacharya, (submitted to Phys. Rev. B) ; http://xxx.lanl.gov/abs/cond-mat/9909297.   
\bibitem{r25} A. B. Pippard, Philos. Mag. {\bf19},  217 (1969); A. M. Campbell and J. E. Evetts, Adv. Phys. {\bf 21}, 327 (1972) and references therein.
\bibitem{r26} A. I. Larkin and Yu. N. Ovchinnikov, J. Low Temp. Phys. {\bf 34},   409 (1979). 
\bibitem{r27} A. I. Larkin, M. C. Marchetti and V. M. Vinokur, Phys. Rev. Lett. {\bf 75}, 2992 (1995); A. Gurevich and V. M. Vinokur, {\bf 83}, 3037 (1999).
\bibitem{r28} M. C. Cha and H. A. Fertig, Phys. Rev. Lett. {\bf 80}, 3851 (1998);  C. Reichhardt, K. Moon, R. Scalettar and G. Zimanyi, ibid. {\bf 83}, 2282 (1999).
\bibitem{r29} P. L. Gammel,  U. Yaron,  A. P. Ramirez,  D. J. Bishop, A. M. Chang, R. Ruel, L. N. Pfeiffer and E. Bucher,  Phys. Rev. Lett. {\bf80}, 833 (1998).
\bibitem{r30} I. Joumard, J. Marcus, T. Klein and R. Cubitt,
 Phys. Rev. Lett. {\bf 82}, 4930 (1999).
\bibitem{r31}C. Bernhard, C. Wenger, Ch. Niedermayer, D. M. Pooke, J. L. Tallon,
Y. Kotaka, J. Shimoyama, K. Kishio, D. R. Noakes, C. E. Stronach, 
T. Sembiring and E. J. Ansaldo, Phys. Rev. B {\bf 52}, R7050 (1995); C. M. Aegerter, S. L. Lee, H. Keller, E. M. Forgan and S. H. Lloyd, 
Phys. Rev. B {\bf 54}, R15661 (1996).
\bibitem{r32} T. V. C. Rao, V. C. Sahni, P. K. Mishra, G. Ravikumar, C. V. Tomy, G. Balakrishnan, D. Mck Paul, C. A. Scott, S. S. Banerjee, N. G. Patil, S. Saha, S. Ramakrishnan, A. K. Grover and S. Bhattacharya, Physica C {\bf 299}, 267 (1998).  
\bibitem{r33} A. I. Larkin and Yu. N. Ovchinnikov, Sov. Phys. JETP {\bf 38}, 854 (1974).
\bibitem{r34} A. I. Larkin, J. Low Temp. Phys. {\bf 34}, 409 (1970).
\bibitem{r35} S. S. Banerjee, N. G. Patil, S. Ramakrishnan, A. K. Grover, S. Bhattacharya, G. Ravikumar, P. K. Mishra, T. V. Chandrasekhar Rao, V. C. Sahni, M. J. Higgins, C. V. Tomy, G. Balakrishnan and D. Mck Paul, Phys. Rev. B, {\bf 59},  6043 (1999).
\bibitem{r36} P. H. Kes, Nature (London), {\bf 376}, 729 (1995); L. A. Angurel, 
 F. Amin, M. Polichetti, J. Aarts and P. H. Kes, Phys. Rev. B  {\bf56}, 3425 (1997).
\bibitem{r37} K. Ghosh, S. Ramakrishnan, A. K. Grover, Gautam I. Menon, Girish Chandra, T. V. Chandrasekhar Rao, G. Ravikumar, P. K. Mishra, V. C. Sahni, C. V. Tomy, G. Balakrishnan, D. Mck Paul and S. Bhattacharya, Phys. Rev. Lett., {\bf 76}, 4600 (1996).
\bibitem{r38} S. S. Banerjee {\it et al}, cond-mat 9911324.
\bibitem{r39} D. R. Nelson, Phys. Rev. Lett. {\bf 60},1973 (1988).
\bibitem{r40} The crystals selected for elucidationg the melting transition in calorimetric data typically have very low levels of quenched impurities.
\bibitem{r41} S. Ramakrishnan, S. Sundaram, R. S. Pandit and G. Chandra, 
J. Phys. E {\bf18}, 650 (1985).
\bibitem{r42} The untwinned crystal in which a giant PE was reported at high fields  in Ref. 13 had 10-90$\%$ width of 0.85K.
\bibitem{r43} C. P. Bean, Rev. Mod. Phys. {\bf 38}, 36 (1964).  
\bibitem{r44} X. S. Ling and J. I. Budnick, in Magnetic Susceptibility of Superconductors and other Spin Systems, edited by R. A. Hein, T. L. Francavilla and D. H. Leibenberg (Plenum Press, New York, 1991), p. 377. 
\bibitem{r45} N. G. Patil {\it et al}, in {\it Advances in Superconductivity:
New Materials, Critical currents and Devices}, Eds. R. P. Pinto {\it et al}
(New Age International Ltd. Publishers, New Delhi, India, 1997), p.335. 
\bibitem{r46} S. S. Banerjee, S. Saha, N. G. Patil, S. Ramakrishnan, A. K. Grover, S. Bhattacharya, G. Ravikumar, P. K. Mishra, T. V. C. Rao, V. C. Sahni, C. V. Tomy, G. Balakrishnan, D. Mck. Paul and M. J.Higgins, Physica C {\bf 308}, 25 (1998).
\bibitem{r47} R. A. Hein and R. A. Falge Jr., Phys. Rev. {\bf 123}, 407 (1961).
\bibitem{r48} N. Morozov, E. Zeldov, D. Majer and M. Konczykowski,
 Phys. Rev. B {\bf 54}, R3784 (1996).
\bibitem{r49}  A. F. Khoder, M. Couach  and J. L. Jorda, Phys. Rev. B {\bf 42},
8714 (1990).
\bibitem{r50} A. K. Grover {\it et al}, Physica C {\bf 220}, 353 (1994);
 Pramana-J. Phys. {\bf 42}, 193 (1994) and Y. Yamaguchi {\it et al}, in {\it Advances in Superconductivity VII}, eds. K. Yamafuji and T. Morishita (Springer-Verlag, Tokyo, Japan, 1995), p. 205.
\bibitem{r51} G. Ravikumar, P. K. Mishra, V. C. Sahni, S. S. Banerjee, S. Ramakrishnan, A. K. Grover, P. L. Gammel, D. J. Bishop, E. Bucher, M. J. Higgins and S. Bhattacharya, Physica C (in press).
\bibitem{r52} S. S. Banerjee, N. G. Patil, S. Ramakrishnan, A. K. Grover, S. Bhattacharya, G. Ravikumar, P. K. Mishra, T. V. Chandrasekhar Rao, V. C. Sahni,
 M. J. Higgins, Appl. Phys. Lett. {\bf 74}, 126 (1999).
\bibitem{r53} P. L. Gammel, D. J. Bishop, G. J. Dolan, J. R. Kwo, C. A. Murray,
L. F. Scneemeyer and J. V. Waszcozak, Phys. Rev. Lett. {\bf 59}, 2593 (1987).
\bibitem{r54} M. V. Marchresky, Ph.D thesis, University of Leiden, 
 The Netherlands (1997).
\bibitem{r55} M. Steingart, A. G. Pute and E. J. Kramer, 
 J. Appl. Phys., {\it 44}, 5580 (1973).
\bibitem{r56} R. Wordenweber,  P. H. Kes and C. C. Tsuei,  
Phys. Rev. B {\bf33}, 3172 (1986).
\bibitem{r57}  W. Henderson, E. Y. Andrei, M. J. Higgins and S. Bhattacharya, Phys. Rev. Lett. {\bf 77}, 2077 (1996).
\bibitem{r58} G. Ravikumar, V. C. Sahni, P. K. Mishra, T. V. C. Rao, S. S. Banerjee, A. K. Grover, S. Ramakrishnan, S. Bhattacharya, M. J. Higgins, E. Yamamoto, Y. Haga, M. Hedo, Y. Inada and Y. Onuki, Phys. Rev. B {\bf57}, R11069 (1998).
\bibitem{r59} S. S. Banerjee {\it et al}, Proc. of First Euro. Conf on Vortex state, Crete, Greece, Sept. 1999, Physica C (in press).
\bibitem{r60} L. Krusin-Elbaum, L.Civale, F. Holtzberg, A. P. Malozemoff and 
C. Field, Phys. Rev. Lett. {\bf 67}, 3156 (1991).
\bibitem{r61} S. Kamal, D. A. Bonn, N. Goldenfeld,
P. J. Hirschfeld, R. Liang and W. N. Hardy, Phys. Rev. Lett. 
{\bf 73}, 1845 (1994).
\bibitem{r62} G. J. Dolan, G. V. Chandrasekhar, T. R. Dinger, C. Feild
and F. Holtzberg, Phys. Rev. Lett. {\bf 62}, 827 (1989).
\bibitem{r63} C. A. Murray, P. L. Gammel, D. J. Bishop, D. B. Mitzi and 
A. Kapitulnik, Phys. Rev, Lett. {\bf 64}, 2312 (1990); D. G. Grier,
C. A. Murray, C. A. Bolle,  P. L. Gammel, D. J. Bishop, D. B. Mitzi and 
A. Kapitulnik, Phys. Rev, Lett. {\bf 66}, 2270 (1992).
\bibitem{r64} M. V. Feigel'man and V. M. Vinokur, Phys. Rev. B {\bf 41}, 8986 (1990).
\bibitem{r65} H. M. Carruzzo and C. C. Yu, Phil. Mag. B {\bf 77}, 1001 (1998).
\bibitem{r66} A. K. Pradhan, S. Shibata, K. Nakao and N. Koshizuka, Europhys. Lett {\bf 46}, 787 (1999).
\bibitem{r67} S. Horiuchi, M. Catoni, M. Uchida, T. Tsurta and
Y. Matsi, App. Phys. Lett. {\bf 73}, 1293 (1998); Bull. Mat. Science,
{\bf 22}, 227 (1999).
\end{references}
\end{document}